\documentclass{article}

\usepackage{arxiv}

\usepackage[utf8]{inputenc} 
\usepackage[T1]{fontenc}    
\usepackage{hyperref}       
\usepackage{url}            
\usepackage{booktabs}       
\usepackage{amsfonts}       
\usepackage{nicefrac}       
\usepackage{microtype}      
\usepackage{lipsum}		
\usepackage{graphicx}
\usepackage{natbib}
\usepackage{doi}
\usepackage{amsmath}
\usepackage{tabularx}
\usepackage{makecell} 
\usepackage{adjustbox}

\title{	SCHEMORA: Schema Matching via Multi-stage Recommendation and Metadata Enrichment using Off-the-Shelf LLMs}

\author{ {\hspace{1mm}Osman Erman  Gungor, PhD. } \\
	Principal Machine Learning Engineer\\
	Informatica\\
	Redwood City, CA, 94063 \\
	\texttt{oermangungor@gmail.com} \\
	\And
	{\hspace{1mm}Derek Paulsen} \\
	Computer Science\\
	University of Wisconsin-Madison\\
	Madison, Wisconsin \\
	\texttt{dpaulsen@informatica.com} \\
        \And
        {\hspace{1mm}William Kang} \\
	Director of AI/ML Development \\
	Informatica\\
	Redwood City, CA, 94063 \\
	\texttt{wkang@informatica.com} \\
}

\date{}

\renewcommand{\shorttitle}{\textit{arXiv} SCHEMORA}

\hypersetup{
pdftitle={A template for the arxiv style},
pdfsubject={q-bio.NC, q-bio.QM},
pdfauthor={David S.~Hippocampus, Elias D.~Striatum},
pdfkeywords={First keyword, Second keyword, More},
}

\begin{document}
\maketitle

\begin{abstract}

Schema matching is essential for integrating heterogeneous data sources and enhancing dataset discovery, yet it remains a complex and resource-intensive problem. We introduce SCHEMORA, a schema matching framework that combines large language models with hybrid retrieval techniques in a prompt-based approach, enabling efficient identification of candidate matches without relying on labeled training data or exhaustive pairwise comparisons. By enriching schema metadata and leveraging both vector-based and lexical retrieval, SCHEMORA improves matching accuracy and scalability. Evaluated on the MIMIC-OMOP benchmark, it establishes new state-of-the-art performance, with gains of 7.49\% in HitRate@5 and 3.75\% in HitRate@3 over previous best results. To our knowledge, this is the first LLM-based schema matching method with an open-source implementation, accompanied by analysis that underscores the critical role of retrieval and provides practical guidance on model selection.

\end{abstract}


\section{Introduction}

Schema matching takes two schemas as input and produces a mapping between their semantically related elements (\cite{rahm2001survey}). It has a wide range of applications in database management and cataloging, such as integrating data from diverse sources, discovering undocumented PK–FK relationships, and identifying joinable tables (\cite{koutras2020valentine}). Additionally, schema matching is vital for developing robust machine learning models by facilitating dataset discovery to enhance feature sets and by validating data to ensure higher training quality (\cite{seedat2024matchmaker}). Yet despite its importance, it often relies on ad-hoc or heuristic approaches, hampered by the lack of standardized benchmarks and rigorous evaluation frameworks that make it difficult to assess progress or compare methods (\cite{koutras2020valentine}).

Manual schema matching remains notoriously time-consuming and error-prone. Organizations report that mapping a single dataset to common standards like OMOP often requires 40--80 hours of manual effort and domain expert review—even for experienced teams (\cite{mecoli2023myositis}). These difficulties are amplified by inconsistent schema designs, incomplete or ambiguous metadata, privacy restrictions that limit access to underlying data, and the sheer scale and imbalance of modern schemas (\cite{zhang2023schema}). As schema sizes grow, manual processes also become more susceptible to cognitive biases and human error (\cite{sheetrit2024rematch}).

To address these challenges, researchers have investigated various approaches, including supervised machine learning techniques (\cite{doan2001lsd}, \cite{he2004}), ensemble and active learning methods (\cite{peukert2011ensemble}, \cite{gal2011}), and a range of deep learning models designed to capture latent semantic relationships via embeddings and neural architectures (\cite{zhang2021smat}, \cite{mudgal2018deepmatcher}, \cite{zhang2023bert}). While these methods can automate schema matching, they also present new drawbacks. Specifically, they often require large amounts of labeled training data, are prone to overfitting within specific domains, and face difficulties adapting to evolving schema conventions—resulting in the need for continuous parameter tuning and retraining to maintain performance (\cite{feng2024promptmatcher}).

Recent advances in large language models (LLMs) offer a promising alternative for building schema matching frameworks without supervised training. \cite{feng2024promptmatcher} proposed a method that computes the cross product of all source and target columns and then uses GPT-4 to perform pairwise verification to identify valid matches. This is followed by complex LLM-based scoring procedures, which add to the already high computational cost of the cross-product approach. For schemas with many tables or columns, such frameworks can quickly become computationally and financially infeasible.

To overcome the inefficiency of exhaustive pairwise comparisons, \cite{sheetrit2024rematch} and \cite{seedat2024matchmaker} propose retrieval-based approaches that bypass the need to evaluate every possible column pair. ReMatch (\cite{sheetrit2024rematch}) serializes column metadata by concatenating elements such as column names, table names, descriptions, and data types into strings, which are then embedded and indexed for semantic retrieval. Similar candidates are retrieved through semantic search and subsequently ranked and filtered by a large language model (LLM). However, as highlighted by \cite{liu2024magneto}, the order in which these elements are concatenated can significantly affect performance. Moreover, embeddings generated from long concatenated strings risk losing critical information (\cite{zhou2024lengthcollapse}). ReMatch is also constrained to 1:1 or m:1 matches, meaning each source column or group of columns can align with only a single target column. Building on ReMatch, Matchmaker (\cite{seedat2024matchmaker}) introduces several improvements, with the primary enhancement being a self-reflection mechanism that retrieves dynamic in-context examples evaluated by another LLM during inference. While this creates a more sophisticated scoring pipeline, it also increases computational demands and inherits ReMatch’s limitations, including the restriction to 1:1 and m:1 mappings and susceptibility to information loss from column serialization.

In this work, we address these challenges by proposing a schema matching framework that leverages off-the-shelf large language models in a purely prompt-driven manner, eliminating the need for annotated data or domain-specific fine-tuning. Rather than serializing raw column metadata, we create enriched representations that capture relevant information, avoiding fragile concatenation schemes that require optimization and tuning. Our approach also removes computationally expensive processes like pairwise comparisons and self-reflection, enabling scalability to large schemas. Through rigorous evaluation on standard benchmarks, we demonstrate that our framework not only addresses the key limitations of prior methods but also achieves new state-of-the-art performance. Furthermore, by including a baseline that performs schema matching without retrieval—relying solely on LLM ranking—we empirically highlight the essential role of retrieval in driving performance. We also identify common pitfalls in existing studies, benchmarks, and evaluation protocols, and extend our evaluation to support both 1:n and m:n matching scenarios, providing broader insights to guide future research.

The main contributions of our work are as follows:

\begin{itemize}
\item We present a schema matching framework that operates entirely without labeled training data, making it highly generalizable and straightforward to deploy across diverse domains and data ecosystems.
\item We introduce the first hybrid retrieval architecture for schema matching, combining vector-based semantic search with BM25 lexical retrieval to capture complementary signals.
\item We demonstrate that metadata enrichment using LLMs is sufficient to align heterogeneous schemas and bridge differences in cross-schema terminology, without relying on complex or computationally intensive methods.
\item On the MIMIC-OMOP benchmark, our framework achieves a new state-of-the-art, improving HitRate@5 by 7.49\% and HitRate@3 by 3.75\% over the previously reported best results.
\item To our knowledge, this is the first LLM-based schema matching framework to openly release its entire codebase enabling transparency, reproducibility, and broader adoption by practitioners and researchers (repo link: https://github.com/ermangungor/schemora).
\item We empirically demonstrate and quantify the critical importance of retrieval for schema matching by comparing against a retrieval-free LLM baseline.
\item We also run extensive experiments to provide practical insights into the effects of LLM size and embedding model choice, offering guidance for machine learning practitioners.
\end{itemize}

\section{Problem Definition}

In this paper, we define the schema matching task as taking two relational schemas as input and producing a mapping between the columns of those schemas that convey the same semantic information.

More formally, we are given a source schema \( S_s \) consisting of a set of tables \( \{T_{s1}, T_{s2}, \ldots\} \), where each table \( T_{si} \) has a set of columns \( \{C_{si1}, C_{si2}, \ldots\} \). Similarly, the target schema is denoted as \( S_t \), consisting of tables \( \{T_{t1}, T_{t2}, \ldots\} \), each with columns \( \{C_{tj1}, C_{tj2}, \ldots\} \). Each table and column is associated with metadata such as a name and a textual description (e.g., \( T_{si}.\text{name} \), \( C_{si1}.\text{description} \)).

The goal of schema matching is to find a mapping function
\[
f:  C_{si1} \rightarrow \mathcal{P}(C_{t})
\]
where \(  C_{si1}\) is a source column and \( C_{t} \) is the set of all target columns, and \( \mathcal{P}(C_{t}) \) denotes the power set of \( C_{t} \). That is, for each individual source column \(  C_{si1}\), the mapping \( f( C_{si1}) \) returns a set of target columns that are semantically equivalent to, or collectively represent, the same information as \( C_{si1}\).

This definition explicitly accommodates both one-to-one and one-to-many correspondences between columns. It also implicitly accommodates many-to-one and many-to-many mappings, since multiple source columns (e.g., \( C_{si1} \) and \( C_{sj2} \)) can be mapped to the same set of target columns.

\section{Methodology}

SCHEMORA consists of two primary steps: indexing and querying. In the indexing step, we prepare the search database by enriching and storing the target columns. The querying step involves finding the matching target column in the database for a given source column. Below, we elaborate on each step.

\subsubsection{Indexing}

Indexing utilizes the target schema to create databases that support vector search and full-text search. This process consists of four steps: enrichment, preprocessing, vector index generation, and full-text search index generation. 

\textbf{Enrichment}: The enrichment step involves two types of prompts designed to enhance schema matching performance. The first prompt guides a Large Language Model (LLM) to expand column names by taking into account the table's name, table description, original column name, and column description. This enrichment aims to transform potentially cryptic column names into semantically rich versions; for instance, "loc\_id" becomes "location identification." While beneficial, this expansion does not fully address inconsistencies in terminology and expressions across different schemas.

To overcome this, a second prompt was developed to instruct the LLM to generate column names without using any words from the table name or column name. This prompt intentionally excludes the column description because it tends to constrain the LLM to the specific schema context, often resulting in the generation of similar names to those produced by the first prompt. By omitting the description, the LLM is encouraged to consider broader semantic possibilities beyond the immediate schema, thus avoiding repetitive naming patterns. For example, the LLM can transform "ward\_id" from a hospital table into "location id," enhancing the likelihood of successful column matches. 

Both prompts utilize the Chain-of-Thought (\cite{wei2022chain}) method with a one-shot example taken from e-commerce, deliberately chosen outside the evaluation domain (i.e., healthcare) to prevent data leakage. We experimented with generating $n$ names per prompt, where $n$ serves as a hyperparameter tested at values of 1,2 and 3. Results are detailed in the Results section.

\textbf{Preprocessing}: In the preprocessing step, we clean the generated names by removing punctuation, replacing all numbers with whitespace, and eliminating all parentheses. Additionally, we split the names using snake case or camel case splitters, since the LLM sometimes generates names like LocationID or location\_id. 

\textbf{Vector Index Generation:} We use FAISS (\cite{faiss}) to create a vector database for vector searches, employing five different embedding models.

\textbf{Full-text Search Index Generation:} We use the BM25 (\cite{bm25s}) Python package, which suits small target schemas, adopting the "lucene" scoring method to maintain compatibility with elastic search. No scoring algorithm tuning was performed, opting instead to optimize other hyperparameters such as the LLM and embedding model.

\subsubsection{Querying}

\textbf{Enrichment and Preprocessing:} These initial steps mirror those introduced in the Indexing section. For brevity, they will not be repeated here.

\textbf{Candidate Retrieval:} Candidates are retrieved using both vector and full-text search for each generated name. Vector search parameters include $top_k=50$ with a cosine similarity threshold of 0.5, while full-text search employs $top_k=50$ with a BM25 score threshold of 1. Although these settings yield high recall, they also introduce reduced precision due to excessive candidate retrieval—addressed in the next step.

\textbf{Table Selection:} A custom prompt directs the LLM to select pertinent tables based on their name and description relative to the source column's table. This is crucial for filtering out common columns, like patient\_id, which may appear in multiple tables, maintaining manageable candidate numbers. 

\textbf{Ranking:} The final ranking step involves compiling all column metadata (original column name, enriched names, table name) and instructing the LLM to rank them. 

\section{Experiment Setup}

\subsection{Baselines}

We selected \cite{sheetrit2024rematch} and \cite{seedat2024matchmaker} as our baselines as they also unsupervised LLM based approaches and also they chose databases which allows mapping columns between multiple tables as opposed to matching a one table to another (such as \cite{liu2024magneto}). Additionally, \cite{seedat2024matchmaker} presents SOTA supervised schema matching tecnique in their paper along with more traditional methods as their baseline which we direclty adopt to our paper.

In addition to these studies, we have developed an intriguing baseline. With the increasing prompt length of LLMs, many practitioners and researchers have begun to question the necessity of retrieval pipelines. The assumption is that if all candidate matches can be included within an LLM prompt, the model should be able to identify the correct match without the need for retrieval. To test this hypothesis, we used an LLM with the entire target schema as a baseline in our experiments.

\subsection{Metric}

Both baseline papers use a metric referred to as \textit{accuracy@K}. The formula for accuracy@K in Equation~\ref{eq:accuracy} is direclty adopted from \cite{sheetrit2024rematch}.

\begin{equation}
\label{eq:accuracy}
\text{accuracy@K} = \frac{1}{N} \sum_{i=1}^{N} \mathbb{I}\left\{ \exists C_{tj} : (C_{si}, C_{tj}) \in \text{match},\ C_{tj} \in f_K(C_{si}) \right\}
\end{equation}
where:
\begin{itemize}
    \item \(N\) is the total number of queries (source columns).
    \item \(f_K(C_{si})\) is the set of top \(K\) candidate matches for source column \(C_{si}\).
    \item \(\text{match}\) is the set of ground truth pairs \((C_{si}, C_{tj})\) indicating correct matches between source and target columns.
    \item \(\mathbb{I}(\cdot)\) is the indicator function that returns 1 if the condition is true, and 0 otherwise.
\end{itemize}

\noindent

\cite{krichene2020sampled} provides a comprehensive overview of evaluation metrics used in the literature. As noted in their work, \textit{accuracy@K} is equivalent to \textit{hitrate@K}, a widely used metric in recommendation systems. Furthermore, when there is only one ground truth target per query, both metrics are equivalent to \textit{recall@K}. The recall@K metric is defined as follows:

\begin{equation}
\label{eq:hit}
\text{Recall@K} = \frac{1}{N} \sum_{i=1}^{N} \frac{\left|\, f_K(C_{si}) \cap \text{GroundTruth}(C_{si})\,\right|}{|\text{GroundTruth}(C_{si})|}
\end{equation}

where:
\begin{itemize}
    \item \(N\) is the total number of queries (source columns).
    \item \(f_K(C_{si})\) denotes the set of top \(K\) candidate target columns returned by the model for the \(i\)-th source column \(C_{si}\).
    \item \(\text{GroundTruth}(C_{si})\) is the set of ground truth target columns for \(C_{si}\).
    \item \(|\cdot|\) denotes set cardinality.
\end{itemize}

\subsection{Data}

We use two datasets that were also employed in our baseline studies. The first dataset, MIMIC, was introduced by the authors of \cite{sheetrit2024rematch}. In this dataset, the schema of MIMIC-III—a public database of deidentified patient records from the Beth Israel Deaconess Medical Center—is manually aligned to the Observational Medical Outcomes Partnership (OMOP) Common Data Model \cite{ohdsi_data_standardization}, an open-source healthcare data standard. This mapping was curated by a domain expert who combined their expertise with an existing mapping from prior work \cite{paris2021transformation}. When no suitable OMOP attribute could be identified, the corresponding MIMIC-III attribute was labeled as NA.  Note that this version differs from the variant available in \cite{JZCS2018_omop_mimic_data}. The full dataset can be accessed at \cite{meniData1_MIMIC_2_OMOP}.

The second dataset, referred to here as Synthea, was introduced by the SMAT study \cite{zhang2021smat}, which proposed a deep learning approach for schema matching. This dataset offers a partial mapping from the Synthea schema—based on a synthetic healthcare dataset (\cite{walonoski2018synthea})—to a subset of OMOP attributes. Unlike MIMIC, which provides a complete mapping of the source schema to OMOP, the Synthea dataset includes only partial correspondences. This dataset is available at \cite{JZCS2018_omop_mimic_data}. Summary statistics for both datasets are presented in Table \ref{tab:dataset_properties}.

\begin{table}[ht]
    \centering
    \caption{Summary of the table properties of our two schema matching datasets.}
    \label{tab:dataset_properties}
    \begin{tabular}{lccc}
        \hline
        Dataset & Source Tables & Target Tables & Mapping Type \\
        \hline
        MIMIC-OMOP & 26 & 14 & 1-to-1 \\
        SYNTHEA-OMOP & 12 & 21 & m-to-n \\
        \hline
    \end{tabular}
\end{table}

A key distinction between these datasets is that MIMIC contains only one-to-one matches, while Synthea includes many-to-many mappings. As discussed in the metric section, the accuracy metric used by prior works is only valid under the assumption of a single ground truth per query (i.e., one-to-one or many-to-one matches). Accordingly, \cite{sheetrit2024rematch} limits their evaluation to one-to-one and many-to-one matches by removing queries with one-to-many correspondences. This restriction leaves only 11 out of 38 query columns, which is insufficient for robust evaluation due to high variance; a few correct predictions by chance can significantly affect aggregate results. The follow-up study \cite{seedat2024matchmaker}, which builds on \cite{sheetrit2024rematch}, presumably applies the same filtering procedure, as it states that it follows the evaluation protocol of \cite{sheetrit2024rematch}, although it does not specify the details. Notably, since neither of these studies open-sourced their code, we can only speculate about the exact filtering and evaluation steps employed. Therefore, we do not report hit rate for this dataset or compare with baselines. Instead, we use the Synthea data to compute recall@k. By reporting these metrics, we aim to enable future work in schema matching to conduct more thorough and meaningful evaluations, facilitating progress toward more robust and reproducible benchmarks in this field.

\section{Results and Discussions}

\subsection{Hyper Parameter Selection}

The selection of language models, embedding models, and the number of generated names are deemed the most critical parameters for SCHEMORA's effectiveness. For language models, the choice between GPT-4.1-2025-04-14 and the smaller GPT-4.1-mini-2025-04-14 was crucial to understanding the impact of model size on performance. Regarding embedding models, we experimented with five distinct options to explore how different embedding strategies affect outcomes. Finally, we varied the number of generated names from one to three to determine how diversity in name generation influences overall success. Specifically, we always generated three names but we pick the first $n$ names depending on the its value. For these evaluations, we used HitRate@5 as our main metric due to its stability as hitrate@1 was to sensitive to small changes. We maintained a consistent temperature setting of zero across all experiments.

The results, detailed in Table \ref{table:search}, show that GPT-4.1 significantly outperformed GPT-4.1-mini, achieving an average increase of approximately 10\% in HitRate@5 on avereage across all embedding models and numbers of generated names. GPT-4.1 demonstrated the capability to generate superior and more diverse names for each column, while more accurately following instructions and ranking candidates.

The best performance was observed when using the text-embedding-3-large embedding model. This is new and updated model from openai (\cite{textembedding3large}). Interestingly, bge-small-en-v1.5 showed somewhat similar performance to large embedding models (other than text-embedding-3-large) illustrating that a smaller model can operate on par with much larger models in certain contexts. This effectiveness is largely because both source and target enriched names typically consist of 1-4 words and lack the rich semantic context found in full sentences and paragraphs that large embedding models are potentially optimized for. Moreover, BGE models are specifically trained to capture both string similarity and semantic similarity using a loss function derived from sparse retrieval like BM25 (\cite{chen2024bge}). 

The lowest HitRate@5 was observed when only one name was generated, likely due to insufficient diversity to capture cross-schema terminologies. In general (8 out of 10 cases), generating three names resulted in a better HitRate@5 compared to two names. Most differences between generating two and three names were marginal, except in a few scenarios where we observed a ~5\% improvement. Based on this trend, we anticipate that generating more than three names will yield only marginal improvements.

\begin{table}[]
\caption{Parameter Search}
\label{table:search}
\resizebox{\textwidth}{!}{%

\begin{tabular}{lllllll} 
\toprule
\textbf{\begin{tabular}[c]{@{}l@{}}Language \\ Model Name\end{tabular}} & \textbf{\begin{tabular}[c]{@{}l@{}}Embedding \\ Model Name\end{tabular}} & \textbf{\begin{tabular}[c]{@{}l@{}}Embedding \\ Dimension\end{tabular}} & \textbf{\begin{tabular}[c]{@{}l@{}}Number of \\ Candidates\end{tabular}} & \textbf{HitRate@1} & \textbf{HitRate@3} & \textbf{HitRate@5} \\ 
\midrule
gpt-4.1-2025-04-14                                                      & all-mpnet-base-v2                                                        & 768                                                                     & 1                                                                        & 41.18\%            & 60.13\%            & 62.75\%            \\ 
gpt-4.1-2025-04-14                                                      & all-mpnet-base-v2                                                        & 768                                                                     & 2                                                                        & 46.41\%            & 67.32\%            & 71.24\%            \\ 
gpt-4.1-2025-04-14                                                      & all-mpnet-base-v2                                                        & 768                                                                     & 3                                                                        & 49.67\%            & 70.59\%            & 76.47\%            \\ 
gpt-4.1-2025-04-14                                                      & bge-small-en-v1.5                                                        & 384                                                                     & 1                                                                        & 43.79\%            & 60.78\%            & 64.71\%            \\ 
gpt-4.1-2025-04-14                                                      & bge-small-en-v1.5                                                        & 384                                                                     & 2                                                                        & 49.67\%            & 70.59\%            & 75.16\%            \\ 
gpt-4.1-2025-04-14                                                      & bge-small-en-v1.5                                                        & 384                                                                     & 3                                                                        & 50.98\%            & 73.20\%            & 77.78\%            \\ 
gpt-4.1-2025-04-14                                                      & text-embedding-3-large                                                   & 3072                                                                    & 1                                                                        & 37.25\%            & 59.48\%            & 64.71\%            \\ 
gpt-4.1-2025-04-14                                                      & text-embedding-3-large                                                   & 3072                                                                    & 2                                                                        & 47.71\%            & 68.63\%            & 74.51\%            \\ 
gpt-4.1-2025-04-14                                                      & text-embedding-3-large                                                   & 3072                                                                    & 3                                                                        & 54.25\%            & 72.55\%            & 80.39\%            \\ 
gpt-4.1-2025-04-14                                                      & text-embedding-3-small                                                   & 1536                                                                    & 1                                                                        & 43.14\%            & 63.40\%            & 68.63\%            \\ 
gpt-4.1-2025-04-14                                                      & text-embedding-3-small                                                   & 1536                                                                    & 2                                                                        & 50.98\%            & 71.90\%            & 77.78\%            \\ 
gpt-4.1-2025-04-14                                                      & text-embedding-3-small                                                   & 1536                                                                    & 3                                                                        & 54.90\%            & 72.55\%            & 78.43\%            \\ 
gpt-4.1-2025-04-14                                                      & text-embedding-ada-002                                                   & 1536                                                                    & 1                                                                        & 45.10\%            & 66.01\%            & 71.90\%            \\ 
gpt-4.1-2025-04-14                                                      & text-embedding-ada-002                                                   & 1536                                                                    & 2                                                                        & 47.06\%            & 70.59\%            & 74.51\%            \\ 
gpt-4.1-2025-04-14                                                      & text-embedding-ada-002                                                   & 1536                                                                    & 3                                                                        & 50.33\%            & 71.90\%            & 76.47\%            \\ 
gpt-4.1-mini-2025-04-14                                                 & all-mpnet-base-v2                                                        & 768                                                                     & 1                                                                        & 35.95\%            & 53.59\%            & 56.21\%            \\ 
gpt-4.1-mini-2025-04-14                                                 & all-mpnet-base-v2                                                        & 768                                                                     & 2                                                                        & 45.10\%            & 64.05\%            & 69.28\%            \\ 
gpt-4.1-mini-2025-04-14                                                 & all-mpnet-base-v2                                                        & 768                                                                     & 3                                                                        & 39.22\%            & 62.75\%            & 67.97\%            \\ 
gpt-4.1-mini-2025-04-14                                                 & bge-small-en-v1.5                                                        & 384                                                                     & 1                                                                        & 28.76\%            & 45.10\%            & 51.63\%            \\ 
gpt-4.1-mini-2025-04-14                                                 & bge-small-en-v1.5                                                        & 384                                                                     & 2                                                                        & 43.79\%            & 61.44\%            & 67.97\%            \\ 
gpt-4.1-mini-2025-04-14                                                 & bge-small-en-v1.5                                                        & 384                                                                     & 3                                                                        & 38.56\%            & 62.75\%            & 67.32\%            \\ 
gpt-4.1-mini-2025-04-14                                                 & text-embedding-3-large                                                   & 3072                                                                    & 1                                                                        & 34.64\%            & 53.59\%            & 56.21\%            \\ 
gpt-4.1-mini-2025-04-14                                                 & text-embedding-3-large                                                   & 3072                                                                    & 2                                                                        & 37.25\%            & 62.09\%            & 67.97\%            \\ 
gpt-4.1-mini-2025-04-14                                                 & text-embedding-3-large                                                   & 3072                                                                    & 3                                                                        & 39.22\%            & 66.01\%            & 69.28\%            \\ 
gpt-4.1-mini-2025-04-14                                                 & text-embedding-3-small                                                   & 1536                                                                    & 1                                                                        & 33.99\%            & 54.90\%            & 62.75\%            \\ 
gpt-4.1-mini-2025-04-14                                                 & text-embedding-3-small                                                   & 1536                                                                    & 2                                                                        & 36.60\%            & 62.75\%            & 68.63\%            \\ 
gpt-4.1-mini-2025-04-14                                                 & text-embedding-3-small                                                   & 1536                                                                    & 3                                                                        & 41.18\%            & 62.09\%            & 69.93\%            \\ 
gpt-4.1-mini-2025-04-14                                                 & text-embedding-ada-002                                                   & 1536                                                                    & 1                                                                        & 31.37\%            & 50.98\%            & 55.56\%            \\ 
gpt-4.1-mini-2025-04-14                                                 & text-embedding-ada-002                                                   & 1536                                                                    & 2                                                                        & 39.87\%            & 60.78\%            & 68.63\%            \\ 
gpt-4.1-mini-2025-04-14                                                 & text-embedding-ada-002                                                   & 1536                                                                    & 3                                                                        & 39.87\%            & 66.01\%            & 69.28\%            \\ 
\bottomrule
\end{tabular}
}
\end{table}

\subsection{Baseline Comparison}

Based on the results in Table \ref{table:search}, we selected the parameter combination with GPT-4.1 as the LLM, \texttt{text-embedding-3-large} as the embedding model, and three generated names per column, as this configuration yields the highest HitRate@5. Using these parameters, we compare our method against existing baselines on the MIMIC dataset, as summarized in Table~\ref{table:baseline}.

\begin{table}[ht]
\centering
\caption{Baseline comparison on MIMIC-OMOP. All results, except for SCHEMORA and Needle-in-the-Stack, are reproduced from \cite{seedat2024matchmaker}.}

\label{table:baseline}
\resizebox{\textwidth}{!}{%
\begin{tabular}{llllllllll}
\toprule
\textbf{Method} 
    & \textbf{\makecell{SCHE\\MORA}} 
    & \textbf{\makecell{Match\\maker}}
    & \textbf{ReMatch} 
    & \textbf{Needle-in-the-Stack} 
    & \textbf{Jellyfish-13b} 
    & \textbf{Jellyfish-7b} 
    & \textbf{LLM-DP} 
    & \textbf{SMAT (20-80)} 
    & \textbf{SMAT (50-50)} \\
\midrule
HitRate@1 & 54.25\% & 62.20\% & 42.50\% & 23.53\% & 15.36\% & 14.25\% & 29.59\% & 6.05\% & 10.85\% \\
HitRate@3 & 72.55\% & 68.80\% & 63.80\% & 45.10\% & N.A. & N.A. & N.A. & N.A. & N.A. \\
HitRate@5 & 80.39\% & 71.10\% & 72.90\% & 62.09\% & N.A. & N.A. & N.A. & N.A. & N.A. \\
\bottomrule
\end{tabular}
}
\end{table}

The results demonstrate that SCHEMORA substantially improves HitRate@3 and HitRate@5 compared to prior methods, outperforming Matchmaker by about 4\% at HitRate@3 (72.05\% vs. 68.8\%) and ReMatch by around 7.5\% at HitRate@5 (80.39\% vs. 72.9\%). This highlights the effectiveness of our approach. Notably, all methods dramatically outperform Needle-in-the-Stack, with margins ranging from 10\% to 20\% depending on the metric.

The Needle-in-the-Stack results are especially instructive: they reveal the necessity of an effective retrieval mechanism. Even if we could technically provide all candidates within the LLM’s context window, the LLM alone struggles to accurately rank correct matches amid a large set of noisy candidates. This underscores the importance of reducing candidate set size—not only to make the problem tractable for LLMs, but also to reduce noise and enable more precise ranking. Our retrieval-based approach directly addresses these challenges and results in substantially improved performance.

\subsubsection{Qualitative Analysis on HitRate@1}

Matchmaker exhibits an unexpected and notably unusual retrieval pattern: although it trails SCHEMORA on HitRate@3 and HitRate@5, and only slightly outperforms ReMatch on these metrics, it achieves a striking 8\% lead in HitRate@1. This result is surprising because models with comparatively lower overall hit rates rarely demonstrate such a pronounced advantage in top-1 accuracy.

Our analysis reveals that Matchmaker frequently succeeds in challenging tie-breaking scenarios, where multiple candidate columns are nearly indistinguishable—often sharing the exact same name but belonging to different tables. A sample of these cases are presented in Table~\ref{table:predictions}. Even for a human reviewer, deciding the correct match in such situations is far from trivial, yet Matchmaker consistently selects the better option.

\begin{table}[ht]
\centering
\caption{Predicted Matches for Source and Target Columns}
\label{table:predictions}
\adjustbox{max width=\textwidth}{%
\begin{tabular}{@{}p{5.5cm} p{5.5cm} m{4.5cm} m{4.5cm}@{}}
\toprule
\textbf{Source} & \textbf{Target} & \textbf{Pred. 1} & \textbf{Pred. 2} \\
\midrule
\makecell[l]{TRANSFERS \\ HADM\_ID}            & \makecell[l]{VISIT\_DETAIL \\ visit\_occurrence\_id}           & \makecell[l]{VISIT\_OCCURRENCE. \\ visit\_occurrence\_id}    & \makecell[l]{VISIT\_DETAIL. \\ visit\_occurrence\_id} \\
\makecell[l]{ICUSTAYS \\ HADM\_ID}             & \makecell[l]{VISIT\_DETAIL \\ visit\_occurrence\_id}           & \makecell[l]{VISIT\_OCCURRENCE. \\ visit\_occurrence\_id}    & \makecell[l]{VISIT\_DETAIL. \\ visit\_occurrence\_id} \\
\makecell[l]{SERVICES \\ HADM\_ID}             & \makecell[l]{VISIT\_DETAIL \\ visit\_occurrence\_id}           & \makecell[l]{VISIT\_OCCURRENCE. \\ visit\_occurrence\_id}    & \makecell[l]{VISIT\_DETAIL. \\ visit\_occurrence\_id} \\
\makecell[l]{OUTPUTEVENTS \\ HADM\_ID}         & \makecell[l]{MEASUREMENT \\ visit\_occurrence\_id}             & \makecell[l]{VISIT\_OCCURRENCE. \\ visit\_occurrence\_id}    & \makecell[l]{MEASUREMENT. \\ visit\_occurrence\_id} \\
\makecell[l]{CALLOUT \\ HADM\_ID}              & \makecell[l]{VISIT\_DETAIL \\ visit\_occurrence\_id}           & \makecell[l]{VISIT\_OCCURRENCE. \\ visit\_occurrence\_id}    & \makecell[l]{VISIT\_DETAIL. \\ visit\_occurrence\_id} \\
\makecell[l]{SERVICES \\ TRANSFERTIME}         & \makecell[l]{VISIT\_DETAIL \\ visit\_detail\_start\_datetime}    & \makecell[l]{VISIT\_DETAIL. \\ visit\_detail\_end\_datetime}   & \makecell[l]{VISIT\_DETAIL. \\ visit\_detail\_start\_datetime} \\
\makecell[l]{ICUSTAYS \\ SUBJECT\_ID}          & \makecell[l]{VISIT\_DETAIL \\ person\_id}                        & \makecell[l]{PERSON. \\ person\_id}                         & \makecell[l]{VISIT\_DETAIL. \\ person\_id} \\
\makecell[l]{OUTPUTEVENTS \\ SUBJECT\_ID}      & \makecell[l]{MEASUREMENT \\ person\_id}                         & \makecell[l]{PERSON. \\ person\_id}                         & \makecell[l]{MEASUREMENT. \\ person\_id} \\
\makecell[l]{CALLOUT \\ SUBJECT\_ID}           & \makecell[l]{VISIT\_DETAIL \\ person\_id}                        & \makecell[l]{PERSON. \\ person\_id}                         & \makecell[l]{VISIT\_DETAIL. \\ person\_id} \\
\makecell[l]{CPTEVENTS \\ HADM\_ID}            & \makecell[l]{PROCEDURE\_OCCURRENCE \\ visit\_occurrence\_id}    & \makecell[l]{OBSERVATION. \\ visit\_occurrence\_id}          & \makecell[l]{PROCEDURE\_OCCURRENCE. \\ visit\_occurrence\_id} \\
\bottomrule
\end{tabular}%
}
\end{table}

What remains unclear is how Matchmaker makes these fine-grained distinctions. One possibility is that it employs a undocumented-particularly effective tie-breaking strategy. Another hypothesis is that the evaluation may be based solely on column names, rather than full (table, column) pairs—an approach that would naturally favor models focused on column-level semantics. However, we cannot verify either explanation, as neither the model implementation nor the evaluation pipeline has been made publicly available.

This lack of transparency presents a fundamental challenge for reproducibility. These findings underscore the importance of open-sourcing both model logic and evaluation frameworks to support meaningful, verifiable progress in schema matching.

\subsection{Ablation Study}

The ablation analysis reveals that all components of the schema matching framework are important for optimal performance, but their contributions vary in magnitude. The removal of query enrichment or document enrichment results in the largest drops across all HitRate metrics, confirming the essential role of semantic information from both the query and document sides. Excluding name expansion or table selection also significantly degrades HitRate@1, emphasizing their impact on accurately identifying the top candidate. By comparison, removing embedding search or full-text search produces only moderate decreases, indicating these retrieval strategies are beneficial but not as critical as enrichment and candidate narrowing. Overall, the results underscore that while every module is valuable, semantic enrichment and table selection are particularly vital for robust schema matching.

\begin{table}[ht]
\centering
\caption{Ablation study on the impact of each component within the schema matching framework. Each row reports performance when the corresponding component is removed.}
\label{table:ablation}
\begin{tabular}{lccc}
\toprule
\textbf{Component Removed} & \textbf{HitRate@1} & \textbf{HitRate@3} & \textbf{HitRate@5} \\
\midrule
Query Enrichment       & 31.58\% & 40.79\% & 42.11\% \\
Document Enrichment    & 32.68\% & 41.83\% & 43.14\% \\
Name Expansion Prompt  & 47.71\% & 63.40\% & 67.32\% \\
Embedding Search       & 53.59\% & 71.90\% & 76.47\% \\
Full-text Search       & 52.29\% & 69.28\% & 76.47\% \\
Table Selection        & 34.64\% & 56.21\% & 68.63\% \\
\bottomrule
\end{tabular}
\end{table}

\subsection{Results for SYNTHEA-OMOP}

As previously discussed, SYNTHEA-OMOP contains many-to-many (m:n) matches, and prior studies have misreported hit rates for this dataset. To establish an accurate baseline for future research, we report recall values for this dataset. Results are presented in Table~\ref{table:recall}.

\begin{table}[ht]
\centering
\caption{Recall performance comparison between SCHEMORA and Needle-in-the-Stack.}
\label{table:recall}
\begin{tabular}{lccc}
\toprule
\textbf{Method} & \textbf{Recall@1} & \textbf{Recall@3} & \textbf{Recall@5} \\
\midrule
SCHEMORA              & 24.33\% & 66.23\% & 80.82\% \\
Needle-in-the-Stack   & 18.10\% & 47.63\% & 59.96\% \\
\bottomrule
\end{tabular}
\end{table}

SCHEMORA consistently outperforms Needle-in-the-Stack across all recall levels, with the gap widening at higher thresholds. These results provide a reliable baseline for future evaluations on SYNTHEA-OMOP.

\section{Related Work}

Pre-machine learning approaches to schema matching were primarily heuristic and structural, relying on manually designed rules to identify correspondences. Rahm and Bernstein~\cite{rahm2001survey} provided a seminal survey that categorized these early efforts into linguistic, structural, and constraint-based methods, laying a comprehensive foundation for the field. Building on such frameworks, Do and Rahm proposed COMA~\cite{do2002coma}, a system that flexibly combined multiple heuristic techniques to improve matching robustness. Similarly, Cupid~\cite{madhavan2001cupid} advanced the state of the art by leveraging hierarchical strategies that integrated schema names, data types, and structural relationships to achieve more precise alignment.

The introduction of traditional machine learning techniques marked a pivotal shift, enabling schema matching systems to learn patterns directly from labeled data. A notable early example is LSD by Doan et al.~\cite{doan2001lsd}, which applied a Naive Bayes classifier to the matching problem. This line of research was extended by He and Chang~\cite{he2004}, who demonstrated how classifiers such as SVMs, decision trees, and early neural networks could significantly improve alignment accuracy. Peukert et al.~\cite{peukert2011ensemble} explored ensemble learning to combine multiple supervised models, while Gal~\cite{gal2011} introduced active learning to reduce annotation costs. YAM++~\cite{bellahsene2011yam} combined supervised classification with heuristic rules for greater adaptability, and other contributions by Duchateau et al.~\cite{duchateau2009} and Berlin and Motro~\cite{berlin2002} highlighted the benefits of multi-level features and decision-tree-driven approaches over purely heuristic baselines.

Embedding models and deep learning architectures have further advanced schema matching by capturing complex latent semantic structures. This category spans from early neural models to more recent transformer-based embedding extraction. Zhang et al.~\cite{zhang2021smat} developed a transformer architecture tailored to learn rich schema representations, while Fernandez et al.~\cite{fernandez2018} leveraged autoencoders, Ebraheem et al.~\cite{ebraheem2018deeper} applied RNNs in DeepER, and Mudgal et al.~\cite{mudgal2018deepmatcher} combined CNNs with attention mechanisms in DeepMatcher to significantly improve performance. Sagi and Gal~\cite{sagi2018} employed Word2Vec embeddings to capture distributional semantics, and Zhang and Balog~\cite{zhang2019} used Siamese networks to compute similarity scores. More recently, transformer models have been fine-tuned or adapted to produce schema embeddings, as demonstrated by Zhang et al.~\cite{zhang2023bert} and Liu et al.~\cite{liu2023smatch}. Magneto~\cite{liu2024magneto} exemplifies this trend by fine-tuning a specialized embedding model for efficient candidate retrieval, which it then combines with large autoregressive models for prompt-based reranking, illustrating how embedding-centric fine-tuning can be tightly integrated with downstream LLM inference.

The field has also explored directly fine-tuning large autoregressive models for schema matching. For example, Jellyfish-7B~\cite{zhang2023b} is an instruction-tuned language model trained across diverse data-centric applications, including schema matching, enabling it to flexibly follow complex alignment directives by modifying its internal parameters.

However, all these training-based methods — whether classical supervised learning, deep embeddings, or fine-tuned large models — fundamentally rely on labeled data, which can be scarce and costly to obtain. They also risk overfitting to training distributions and may suffer from model or data drift when underlying schema conventions evolve. In contrast, a growing class of approaches sidesteps these challenges by employing large autoregressive models purely through prompt engineering and retrieval augmentation, without any additional fine-tuning or supervised data. Sheetrit et al.~\cite{sheetrit2024rematch} introduced a retrieval-enhanced framework that utilizes GPT to generate schema matching predictions on the fly, while Seedat and van der Schaar~\cite{seedat2024matchmaker} proposed a self-improving approach that dynamically refines prompts based on prior outputs. PromptMatcher~\cite{feng2024promptmatcher} demonstrated how carefully crafted GPT-4 prompts could reduce uncertainty in matching results, highlighting the effectiveness of prompt-centric techniques that rely solely on in-context learning. Our work falls within this category, leveraging prompt-based schema matching to avoid the limitations of training-intensive approaches while achieving robust semantic alignment across heterogeneous schemas.

\section{Limitations and Future Work}

While SCHEMORA demonstrates strong performance, generating a large number of enriched names per column can lead to increased index storage requirements. To address this, future work could explore more efficient indexing strategies. For example, in the case of BM25, concatenating all enriched names for a column into a single document may significantly reduce index size, although this approach might affect token frequency statistics. For vector-based indexes, pooling techniques such as mean or max pooling could be employed to create a single representative embedding per column, avoiding the need to store embeddings for each enriched name individually. Alternatively, clustering or filtering methods could be used to retain a diverse subset of enriched names and embeddings, thereby reducing redundancy while preserving semantic richness. Investigating these approaches could improve index storage efficiency and scalability, enabling SCHEMORA to be applied to larger and more complex datasets.

\section{Summary and Conclusions}

This paper introduced SCHEMORA, a schema matching framework that leverages off-the-shelf large language models and prompt-driven metadata enrichment to align heterogeneous schemas without requiring any labeled data, supervised training, or fine-tuning. By generating context-aware column names and combining semantic and lexical retrieval, SCHEMORA effectively addresses inconsistencies in cross-schema terminology. Our experiments on public healthcare benchmarks demonstrate that SCHEMORA achieves new state-of-the-art performance, improving HitRate@5 on MIMIC-OMOP by 7.5\% over previous methods. Notably, we show that it is possible to reach state-of-the-art results using LLMs without any supervised training or domain-specific fine-tuning. Ablation studies and qualitative analysis further highlight that metadata enrichment and multi-stage retrieval are essential for robust schema matching.

Our findings clearly show that metadata enrichment not only enhances schema alignment but can play a significant role in broader database management tasks. The effectiveness of our approach underscores the value of integrating LLM-driven enrichment with retrieval techniques to overcome traditional limitations in schema matching.

Looking ahead, future work will focus on reducing the increased index storage and computational costs introduced by generating multiple enriched names per column. Promising directions include concatenating enriched names into a single document for BM25 indexing, applying pooling strategies for vector embeddings, and filtering for diverse enriched names to minimize redundancy. These efforts aim to further optimize index efficiency and extend SCHEMORA’s applicability to larger and more complex schema matching challenges.


\bibliographystyle{unsrtnat}
\bibliography{SCHEMORA}  






\end{document}